\documentclass[referee]{aa}

\usepackage{graphicx}
\usepackage{natbib}
\usepackage{latexsym}
\usepackage{amsmath}
\usepackage{amssymb}
\usepackage{amstext}
\usepackage{color}
\usepackage{psfrag}
\usepackage{txfonts}

\def\k{km s$^{-1}$}

\def\d{$^\circ$}
\def\m{$^\prime$}
\def\s{$^{\prime\prime}$}

\def\cm3{cm$^{-3}$}

\def\2{$^{12}$CO}
\def\3{$^{13}$CO}
\def\msol{M$_\odot$}
\def\msolyr{M$_{\odot}/$yr}

\begin{document}

\title{The molecular gas around the luminous blue variable star G24.73+0.69}
\author {A. Petriella \inst{1,2}
\and S. A. Paron \inst{1,3}
\and E. B. Giacani  \inst{1,3}
}

\institute{Instituto de Astronom\'{\i}a y F\'{\i}sica del Espacio (CONICET-UBA),
             CC 67, Suc. 28, 1428 Buenos Aires, Argentina\\
            \email{apetriella@iafe.uba.ar}
\and CBC - Universidad de Buenos Aires, Argentina
\and FADU - Universidad de Buenos Aires, Argentina
}

\offprints{A. Petriella}

   \date{Received <date>; Accepted <date>}

\abstract
{}
{We study the molecular environment of the luminous blue variable star G24.73+0.69 to 
investigate the origin of the two infrared shells around this massive star and 
determine its effects on the surrounding interstellar medium.}
{We analyze the distribution of the molecular gas using the \3 J=1--0 emission extracted from 
the Galactic Ring Survey. We use near and mid-infrared data from 2MASS and GLIMPSE to identify 
the young stellar objects in the field.}
{We discover the molecular counterpart to the outer infrared shell around G24.73+0.69. 
The CO shell was probably blown by the stellar wind
of the star mainly during its main sequence phase. 
We also find molecular gas that corresponds to the inner infrared shell, 
although its origin remains uncertain. 
We identify seven young stellar objects
within the molecular material, whose birth might have been triggered by the stellar wind of the luminous blue
variable star.
We suggest that both G24.73+0.69 and the progenitor of the nearby supernova remnant G24.7+0.6 
were formed from the same natal cloud and represent the most evolved members of 
a so far undetected cluster of massive stars.}
{}

\titlerunning{The molecular gas around G24.73+0.69}
\authorrunning{A. Petriella et al.}


\maketitle

\section{Introduction}
\label{secc_pres}

Luminous blue variable (LBV) stars are very massive objects that evolve
from O-type main sequence (MS) stars burning hydrogen in their core to become Wolf-Rayet (WR) 
helium core burning stars \citep{maeder89, hump89}. This transitional stage is 
characterized by a high mass-loss rate (typically between $10^{-5}$ and $10^{-4}$~\msolyr)
sometimes accompanied by so-called giant eruptions, and significant photometric variability
on timescales from months to years \citep{hump94}. The LBV phase is short-lived and as a 
consequence only 12 cases and 23 candidates have been reported in our Galaxy so far \citep{clark05}.
As a result of the mass loss, most of the LBV stars (either confirmed or candidates) are surrounded by a 
nebula that expands with characteristic velocities of between 30 \k and 200 \k (see \citealt{clark05}). These nebulae
have dusty and gaseous components (of a few solar masses) and a large variety of morphologies, from
circular to bipolar \citep{nota95,weis01}. 

Since nebulae around LBV stars are strong emitters in the infrared (IR) regime, most of them 
have been observed in this wavelength range. 
At present, there are only a few cases in which the chemical evolution 
and the presence of circumstellar molecular gas related to these nebulae have been 
determined by molecular studies, the most representative sources
being AG Car \citep{nota02} and G79.29+0.46 \citep{rizzo08,jime10}.
Molecular material was also discovered around the yellow hypergiant (YHG) stars IRC+10420 and 
AFGL2343* \citep{carrizo07}. Theoretical models of stellar evolution and 
observations of post-MS massive stars \citep{meynet00,hump94} suggest that low luminous LBVs 
may pass through a cooler red supergiant (RSG)/yellow hypergiant phase prior to the LBV period.
The RSG/YHG phase is also characterized by mass-loss activity (for a review, see \citealt{dejager98})
that can lead to the presence of molecular gas around these massive stars.   
By analyzing the molecular material, we can determine the mass-loss history of the 
massive star and its effects on the surrounding interstellar medium (ISM).

\section{Presentation of G24.73+0.69}
\label{present}

\begin{figure}[h]
\centering
\includegraphics[width=10cm]{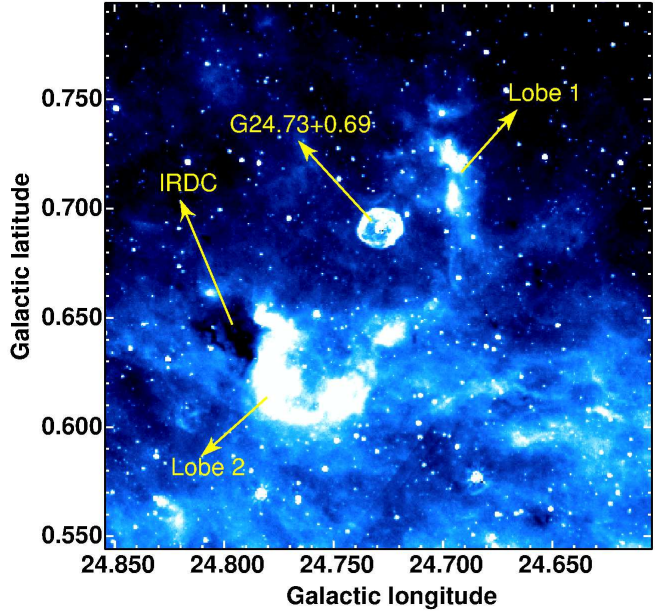}
\caption{Emission in the {\it Spitzer}-IRAC 8 $\mu$m band toward the LBV star G24.73+0.69. 
We indicate the positions of the LBV star and its nebula,
the lobes of the bipolar outer shell, and the IRDC 024.789+0.633. }
\label{fig1}
\end{figure}

The ring nebula around the LBV star G24.73+0.69 (hereafter G24) was discovered by \citet{clark03} 
in the mid-IR. Fig. \ref{fig1} displays the emission in the {\it Spitzer}-IRAC 8 $\mu$m band toward G24,
which clearly shows the nebula around the central massive LBV star.
It has a roughly circular morphology 
with a slight elongation in a direction parallel to the Galactic plane.
Its shape can be approximated by an ellipse of 40\s$\times34$\s.
Taking an expansion velocity of 200 \k,
\citet{clark03} estimated a mass-loss rate
of $\sim9.5\times10^{-5}$\msolyr~for the LBV star and an age of $\sim4,800$ yr for the nebula.
They quoted an upper limit of 5.2 kpc for the distance based on the consistency between the modeled and
the observed spectral energy distribution of the star.
In addition, there is a fragmentary outer shell (OS) of a bipolar morphology with
lobes (labeled 1 and 2 in Fig. \ref{fig1}) projecting into opposite directions.
This structure forms an incomplete shell of 4.6\m$\times2.3$\m~oriented 
$\sim60$\d~with respect to the Galactic plane, whose geometric center 
does not match the position of the LBV star. 
Fig. \ref{fig1} also shows the position of the IR dark cloud (IRDC) 024.789+0.633, 
which is adjacent to the lobe 2. All IRDCs are considered as the precursors of
protostars, being a signature of active star formation \citep{rath06,rath07}.

To explain the origin of the IR shells, \citet{clark03} argued that the inner one
(i.e. the ring nebula) originates in material ejected from the central star during the LBV phase.
However, based on the size and age of the bipolar OS, they 
suggest that it formed from the interaction between the stellar 
wind and the ISM and/or the natal molecular cloud (MC) in a period 
prior to the current phase. 

In this work, we study the molecular gas toward G24.73+0.69 and its environment 
with the aim of unveiling the origin of these IR shells. 
  
\section{Results and discussion}
\label{secc_molec}

\subsection{The distribution of the molecular gas}
\label{distrib}

We study the molecular gas around G24 using data from the Galactic Ring Survey (GRS, \citealt{jackson06}).
The GRS was performed by the Boston University and the
Five College Radio Astronomy Observatory (FCRAO). The survey maps the Galactic ring in the \3 J=1--0 line
with an angular and spectral resolutions of 46\s~and 0.2 \k, respectively.
The observations were performed in both position-switching and on-the-fly mapping modes, achieving an
angular sampling of 22\s.

\begin{figure*}[tt]
\centering
\includegraphics[width=13cm]{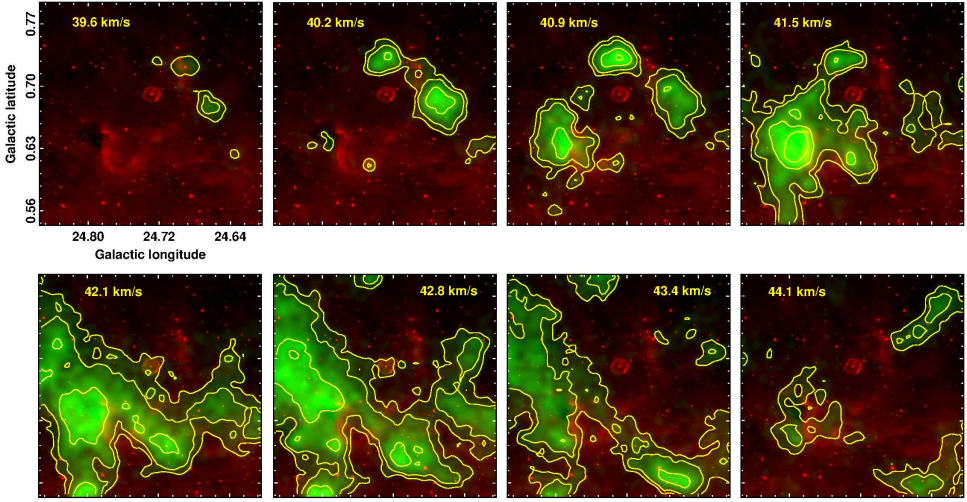}
\caption{Two-color image of the field around the LBV star. In red: {\it Spitzer}-IRAC 8 $\mu$m band.
In green: the integrated \3 J=1--0 emission every 0.7 \k. The CO contours levels are 0.5, 0.9, 1.8, and 2.7 K \k.}
\label{fig2}
\end{figure*}

From the inspection of the \3 data cube in the whole velocity range, we found 
morphological signatures of a possible association between G24 and the surrounding molecular material in 
the velocity range between +39 and +44 \k (all velocities here being referred to the local standard of rest). 
In Fig. \ref{fig2}, we show the 
integrated velocity channel maps of the \3 J=1--0 emission every 0.7 \k~with 
the mid-IR emission from the {\it Spitzer}-IRAC 8 $\mu$m band. 
From this figure, we can discern several condensations of molecular gas forming a molecular shell 
around G24 (see mainly panels at +40.9 and +41.5 \k).
This shell delineates the exterior border of the bipolar OS.
This distribution indicates that there is a connection between the molecular gas and the bipolar OS.
If this were the case, both of them would be at the same distance.
Using the rotational model of \citet{fich89} and adopting a velocity of +42 \k~as 
a systemic velocity for the molecular shell, we obtained kinematic distances of either 3.5 kpc or 12 kpc.
Lacking any additional discriminator between the two kinematic distances and 
taking into account that a distance of 5.2 kpc is quoted as an upper limit for G24 \citep{clark03},
we adopted a distance of 3.5 kpc as the most plausible value for the LBV star and the molecular shell.
This new distance led to a substantial reduction in the calculated intrinsic luminosity of G24. 
\citet{clark03} quoted $log(L_{\star}/L_{\odot})=5.6$ for a distance of 5.2 kpc. Taking a distance
of 3.5 kpc, the luminosity drops to $log(L_{\star}/L_{\odot})\sim5.25$, which would make it one 
of the faintest LBVs yet identified (see for comparison the HR diagram of \citealt{clark09}), thus 
a possible post RSG/YHG object.

We were able to derive a number of parameters characterizing the CO. 
By assuming local thermodynamic equilibrium (LTE), we estimated the H$_{2}$ mass of the molecular shell using 
the equations
$${\rm N(^{13}CO)} = 2.42 \times 10^{14} \frac{T_{\rm ex}+0.88}{1-e^{-5.29/T_{\rm ex}}} \int{\tau_{13} dv},$$
where $T_{\rm ex}$ is the excitation temperature of the \3 transition and $\tau_{13}$ is the optical depth of the line.
Assuming that the \3 J=1--0 line is optically thin, we can use the approximation
$$\int{\tau_{13} dv} \sim \frac{1}{J(T_{\rm ex}) - J(T_{\rm b})} \int{T_{\rm B} dv}, $$
where $$J(T)=\frac{5.29}{e^{5.29/T}-1}, $$ $T_{b}=2.7$~K is the background temperature, 
and $T_{B}$ is the brightness temperature of the line.
We assumed that $T_{ex}=10$~K and used the relation N(H$_{2}$)/N(\3)$ \sim 5 \times 10^5$ \citep{simon01}
to estimate the column density of molecular hydrogen N(H$_{2}$).
The mass of the molecular clumps was calculated from 
$$ {\rm M} = \mu~m_{{\rm H}} \sum{\left[ D^{2}~\Omega~{\rm N(H_{2})} \right] }, $$
where $\Omega$ is the solid angle subtended by the \3 J=1--0 beam size, $m_{\rm H}$ is the hydrogen mass,
$\mu$ is the mean molecular weight assumed to be 2.8 by taking into account a relative helium abundance
of 25 \%, and $D\sim3.5$~kpc is the distance. 
Our summation was performed over the area of each molecular clump.
We summed the mass of the clumps around G24 and 
found that the molecular shell has a total mass of $\sim1,000$~\msol.
The original ambient density $n_{0}$, an important parameter to study the dynamics of the shell, 
could be estimated by assuming that the calculated total mass was originally distributed
in the volume limited by the bipolar OS, which has a mean radius of $\sim3.5$\m. At a distance 
of 3.5 kpc, this corresponds to a mean radius $r_{0}\sim3.5$~pc. Using this value, 
we found that $n_{0}\sim230$~cm$^{-3}$.

\subsection{The origin of the molecular shell}
\label{origin}

There is compelling evidence of shells of interstellar material 
produced by the effect of the strong stellar winds of massive stars on the ISM. 
These so-called stellar wind bubbles have been observed in neutral \citep{cappa05,cicho08,giacani11} 
and molecular \citep{cappa09, cappa10} gas.
We explored the hypothesis that the action of the stellar wind of the LBV star and/or
its progenitor O-type star has blown a cavity of interstellar material piling up the molecular gas 
in the border and forming the molecular shell discovered in the present work.
 
To test this scenario, we compared the mechanical energy $E_{\omega}$
released by the star into the ISM during the lifetime of the molecular shell,
with the kinetic energy $E_{k}$ of the swept-up material. We first 
estimated the dynamical age ($t_{dyn}$) of the shell using the 
equation of \citet{whitworth94b} for a stellar wind bubble:
$t_{dyn}\sim0.02r_{0}^{5/3}L_{37}^{-1/3}n_{3}^{1/3}$~Myr, where $r_{0}$ is
the radius of the shell in pc, $L_{37}$ is the mechanical luminosity $L_{\omega}$
of the stellar wind in units of ${10^{37}}$~erg s$^{-1}$, and $n_{3}$ is the original ambient density in
units of $10^{3}$~\cm3. We calculated $L_{\omega}=1/2\dot{M}v_{\omega}^{2}$ by taking the 
typical parameters of an O-type star in \citet{prinja90} and \citet{mokiem07}, namely a
mass-loss rate $\dot{M}=2\times{10^{-6}}$~$M_{\odot}/yr$ and a stellar wind velocity $v_{\omega}=2,000$~\k.
Thus, we found that $L_{\omega}\sim2.5\times{10^{36}}$~erg s$^{-1}$.
Taking the mean radius of the shell $\sim3.5$~pc and the original ambient density
$\sim230$~\cm3, we evaluated the $t_{dyn}$ to be $\sim0.2$~Myr.
From this result, we can see that the CO shell is more than an order of magnitude older than 
the age of the LBV phase, which typically has a duration of $\lesssim10^{4}$~yr. 
This indicates that the detected CO shell may be the material swept-up by
the stellar wind of the central massive stars during its MS phase.

The kinetic energy can be estimated as $E_{k}=1/2Mv_{e}^{2}$,
where $M$ is the mass of the shell and $v_{e}$ its expansion velocity.
We calculated the expansion velocity to be 
$v_{e}\sim6.2L_{37}^{1/5}n_{3}^{-1/5}t_{dyn}^{-2/5}$~\k~\citep{whitworth94b}, where $t_{dyn}=0.2$~Myr. 
We obtained $v_{e}\sim13$~\k. Using $M=1,000$~M$_{\odot}$, we inferred that
$E_{k}\sim1.7\times{10^{48}}$~erg. For the mechanical energy, we assume that 
$E_{\omega}=L_{\omega}t_{dyn}\sim1.6\times{10^{49}}$~erg.
We calculated the efficiency of the energy conversion of the wind $\epsilon=\frac{E_{k}}{E_\omega}$ and
obtained a value of $\sim0.11$. This agrees with the expected value 
for an energy-conserving stellar wind bubble \citep{mellema02}, hence we conclude that the central star 
has blown a stellar wind strong enough to create the molecular shell.
We also evaluated the contribution of the star during the LBV phase to the energy outflow. 
Taking $\dot{M}=9.5\times{10^{-5}}$~$M_{\odot}/yr$ and $v_{\omega}=200$~\k~ from \citet{clark03} and
$\sim10^{4}$~yr as a typical duration for the LBV phase,
we obtained $E_{\omega}^{LBV}\sim3.8\times{10^{47}}$~erg, which represents less than 3 \% of the energy 
released by the star during the MS phase.

These calculations link the formation of the molecular shell 
to the stellar wind of the central star during the MS phase, 
in good agreement with the origin of the bipolar OS suggested by \citet{clark03}.  
In this scenario, the elongated morphology of the 
bipolar OS (and also of the molecular shell) 
may be a consequence of the asymmetric stellar wind from the central star.
Evidence of non-spherical winds has been found in other massive stars.
For example, \citet{meaburn04} discovered a giant lobe projecting from the ring nebula around 
the LBV star P Cygni. According to the authors, the lobe formed from the mass-ejection activity
prior to the current eruptions that produced the nebula.
They suggested that the peculiar morphology of this giant lobe is due to
the interaction between the ejected material and an asymmetric cavity in the ISM blown by the toroidal stellar
wind of the central star during the MS phase.

Finally, we note the presence of molecular gas superimposed on the ring nebula (see Fig. \ref{fig2}, panels at +42.1 and +42.8 \k).
This material may i$)$ originate in a mass-ejection event from the central
LBV star, as observed in the ring nebula around G79.29+0.46 \citep{jime10}; 
or ii$)$ be part of the molecular shell. Using the equations of Sect. \ref{distrib}, we estimated 
a mass of $\sim20$~\msol~and a density of $\sim120$~cm$^{-3}$. The obtained mass is higher than the mass of the nebula around 
G24 estimated by \citet{clark03} ($\sim0.45$~\msol) and of the nebulae around other LBVs (see \citealt{smith06} and \citealt{clark09}).
Thus, we propose that the molecular material located right upon the ring nebula belongs mainly to the molecular shell.
However, we cannot rule out the presence of some molecular gas ejected by the LBV star. 
We plan to observe a field around G24 with higher density tracers and higher resolution to confirm
the presence of this material.

\subsection{Star formation activity around G24.73+0.69}
\label{secc_starform}

The environment of the LBV star G24 might be one that promotes triggered star formation. 
As we have shown in the previous section,
the massive star has been interacting with the neighboring molecular gas by means of its strong stellar wind.
To search for primary tracers of star formation,
we identified the young stellar object (YSO) candidates using the GLIMPSE Point Source Catalog
(GPSC) in the {\it Spitzer}-IRAC bands.
We constructed a [5.8]-[8.0] versus [3.6]-[4.5] color-color (CC) diagram for the sources
around G24. In the CC diagram, we used the photometric criteria of \citet{allen04} to
classify the sources according to their evolutionary stage: class I are protostars with prominent circumstellar envelopes,
class II are disk-dominated objects, and class III are the most evolved YSOs whose emission comes
mainly from the central star.

\begin{figure}[h]
\centering
\includegraphics[width=8cm]{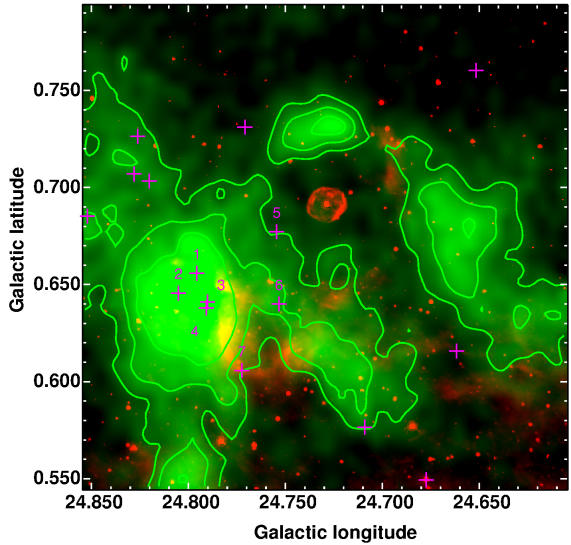}
\caption{Two-color image toward G24.73+0.69. In green with contours: emission of the \3 integrated between +39 and +44 \k
(contour levels are 2.7, 4.4, and 6.2 K \k).
In red: 8 $\mu$m band. The magenta crosses are the class I YSO candidates selected according to the criterion discussed
in the text.}
\label{fig3}
\end{figure}

Fig. \ref{fig3} displays the distribution of the class I YSO candidates around G24.
We can discern seven of these YSO candidates superimposed on the molecular gas swept-up by
the massive star. Four of them (sources 1, 2, 3, and 4) spatially coincide with the maximum of molecular emission.
We note that this region coincides with the IRDC 024.789+0.633 (indicative of active star-forming activity)
and lies next to the lobe 2 of the bipolar OS.

We fit the spectral energy distribution (SED) of the class I YSO
candidates using the fluxes in the 2MASS J, H, and Ks bands and in the four {\it Spitzer}-IRAC bands.
We used the tool developed by \citet{robit07} and available
online\footnote{http://caravan.astro.wisc.edu/protostars/}. We assumed the distance to be between 3 and 4 kpc.
In Table \ref{table2}, we report the class I YSO candidates
magnitudes in the 2MASS and {\it Spitzer}-IRAC bands.
\citet{robit06} defined three evolutionary stages based on the values of the central source mass $M_\star$,
the disk mass $M_{disk}$,
the envelope mass $M_{env}$, and the envelope accretion rate $\dot{M}_{env}$.
Stage I YSOs are those that have $\dot{M}_{env}/M_\star>10^{-6}~yr^{-1}$, i.e., protostars with
large accretion envelopes;
stage II are those with $\dot{M}_{disk}/M_\star>10^{-6}~yr^{-1}$ and $\dot{M}_{env}/M_\star<10^{-6}~yr^{-1}$, i.e.,
young objects with prominent disks; and stage III are those with $\dot{M}_{disk}/M_\star<10^{-6}~yr^{-1}$ and
$\dot{M}_{env}/M_\star<10^{-6}~yr^{-1}$, i.e., evolved sources where the flux is dominated by the central source.
The goodness of the fitting for each model was measured by a $\chi^{2}$ value.
We defined the ``selected models" as those satisfying the equation $\chi^{2}-\chi^{2}_{min}<2N$, where
$\chi^{2}_{min}$ corresponds to the best-fitting model and
{\it N} is the number of input data fluxes (fluxes specified as upper limit do not contribute to {\it N}).
The fitting tool also fit the IR fluxes to a stellar photosphere to check whether the source
could be modeled by a main-sequence star with interstellar reddening. The goodness of the fitting is defined
by a $\chi^{2}_{star}$ value.

\begin{table*}[h]
\caption{2MASS and {\it Spitzer}-IRAC magnitudes of class I YSO candidates.}
\label{table2}
\tiny
\centering
\begin{tabular}{cccccccccc}
\hline\hline
Source & Source (GPSC)    & 2MASS Qual. & $J$ & $H$ & $K_{S}$ &  3.6 $\mu$m &
4.5 $\mu$m  &  5.8 $\mu$m &  8.0 $\mu$m \\
\hline
1     & G024.7956+00.6558 & N/D &        &        &        & 13.047 &   12.173& 11.715 & 11.321 \\
2     & G024.8051+00.6456 & UUA & 18.131  & 17.059 & 13.707 & 10.548 &  9.589 & 8.791  & 8.059 \\
3     & G024.7901+00.6410 & N/D &         &        &        & 12.874& 11.096&   9.983 & 9.639 \\
4     & G024.7907+00.6379 & N/D &         &        &        &13.439&    12.518& 11.638& 11.059 \\
5     & G024.7545+00.6771 & AAA & 14.818 &      13.227 &        12.158  & 10.967&       10.497& 10.194& 9.859\\
6     & G024.7532+00.6400 & UUA & 14.643&       12.784& 12.312  &       11.792& 11.463& 11.722& 11.386 \\
7     & G024.7735+00.6068 & AAA &  13.898       &13.172 &12.803 &11.907&        11.485& 10.083& 8.577\\
\hline
\end{tabular}
\tablefoot{2MASS Qual.: A is the best photometric quality, with a snr $\geq10$, and U means that the magnitude
value is an upper limit. N/D indicates no detection of the source in the 2MASS survey.}
\end{table*}

In Table \ref{table3}, we report the results of the SED fitting for the seven class I candidates
within the molecular shell.
For all of them, all the selected models correspond
to stage I and II (except source 2 for which we get one stage III model).
Moreover, for all of them apart from source 6 we get $\chi^{2}_{min}<\chi^{2}_{star}$. Thus, we can
confirm that they are young sources at the earlier stages of evolution.

\begin{table}[h]
\caption{Results of the SED fitting for class I YSO candidates. In the last column,
we report the evolutionary stage of the source, based on the criterion of \citet{robit06} for the selected models.
When the models correspond to different stages, the less likely is presented in brackets.}
\label{table3}
\centering
\tiny
\begin{tabular}{cccc}
\hline\hline
Source & $\chi^{2}_{min}/{\it N}$ & $\chi^{2}_{star}/{\it N}$ & Stage  \\
       &                           &                            & \\
\hline
1     &  0.4        &     15                    & I or (II) \\
2     &  4          &     77                    & II        \\
3     & 0.1        &     93                     & I         \\
4     & 0.02       &     37                     & I or (II) \\
5     & 0.9        &     4                      & II or (I) \\
6     &  2.3        &     2                     & II        \\
7     & 10         &     91                     & II        \\
                                                                                                                             
\hline
\end{tabular}
\end{table}

As we note in the previous section, the molecular shell might have formed
as a consequence of the interaction between the stellar wind of the central massive star and the
ISM. The presence of YSO candidates within the molecular shell and an IRDC are signs of
active star-forming activity. We suggest that the birth of
these new stars might have been triggered by the expanding wind bubble acting on the
molecular gas.

\subsection{G24.73+0.69 and its surrounding: the big picture}
\label{global}

Massive stars form in giant molecular clouds which, under certain circumstances, 
fragment into dense clumps that collapse to form new stars. 
These stars spend a short period 
of time ($\sim10^{6}$~yrs) in the MS before evolving to become LBV and WR stars, which 
eventually end their lifetimes exploding as supernovae.
These explosions usually occur when the progenitor star still lies close to its natal cloud and 
while other companion stars may still be in the MS or LBV/WR phase.
As a consequence, several supernova remnants have been observed interacting with evolved 
massive stars (see \citealt{velaz03,sush11}) and molecular gas (see \citealt{jiang10} for an exhaustive list
of SNR/MC associations).  

\begin{figure}[h]
\centering
\includegraphics[width=8cm]{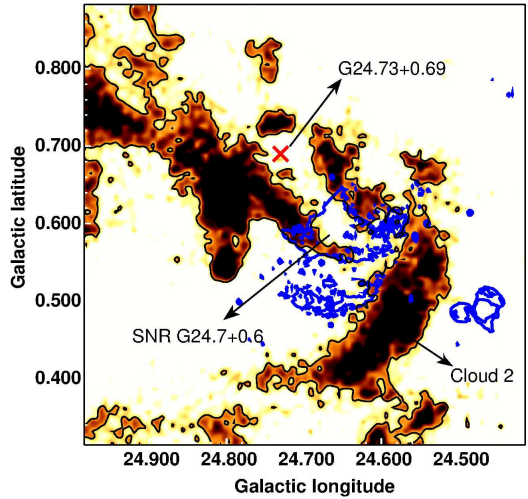}
\caption{Emission of the \3 integrated between +38 and +50 \k. The blue contours are the
radiocontinuum emission at 20 cm of the SNR G24.7+0.6 taken from the MAGPIS. 
The red cross indicates the position of the LBV star G24.73+0.69.}
\label{fig4}
\end{figure}

The LBV star G24.73+0.69 lies $\sim7$\m~from the SNR G24.7+0.6.
In the radio band, this remnant displays a couple of incomplete shells and a 
polarized filled central core with a flat spectrum \citep{reich84},
which indicates that it consists of a plerion powered by an undetected pulsar.
Thus, the SNR formed from the collapse of a massive star.
\citet{petriella08} reported the existence of molecular gas interacting 
with G24.7+0.6 in the velocity range between +38 and +50 \k, which interestingly places the SNR 
at the same kinematic distance as G24. In addition, active star formation probably 
triggered by the SNR and/or its progenitor was also discovered around the remnant \citep{petriella10}.
In Fig. \ref{fig4}, we show the emission of the \3 J=1--0 integrated in the velocity range between +38 and +50 \k.
We indicate the position of the different features in the field: the LBV star (red cross), 
the SNR (in blue contours), and the MC interacting with
the remnant (cloud 2 in the nomenclature of \citealt{petriella08}).
                                                                                                                             
From the previous figure, we see that this large region, which harbors a rich variety of objects 
(a LBV star and its nebula, a SNR, abundant molecular material hosting dark clouds, possible  
star-forming sites), is an ideal laboratory to investigate the genetic connection between 
different stellar populations. On the basis of the morphology and distribution of the molecular gas, it is 
likely that the progenitor star of the SNR G24.7+0.6 and the LBV star may have formed from the same giant 
MC that now we observe fragmented into smaller features. 
They may be the most evolved members of a so far undetected cluster/association of massive stars that also formed
from the same molecular material. In addition, they may have triggered the formation of a second generation of 
stars that now we observe as young objects deeply embedded in the molecular gas. 
This scenario should be investigated further by searching for the missing intermediate-age stellar population, 
namely massive stars still in the MS phase, which formed together with the LBV and the supernova progenitor. 

\section{Summary}
\label{secc_summ}

We have analyzed the \3 emission in the surroundings of the LBV star G24.73+0.69.
We have discovered a fragmented molecular shell in the velocity range between +39 \k and +44 \k, which delineates
the infrared bipolar outer shell of G24. 
On the basis of spatial, morphological, and kinematic coincidence of features, we suggest that there is a 
connection between this IR shell and the molecular gas. 
We argue that the molecular shell formed from the interstellar material swept-up by 
the stellar wind of the central star mainly during its MS phase.
The elongated morphology of the bipolar and molecular shells may be a consequence of the asymmetric stellar wind 
from the central massive star.
We have also detected molecular emission probably associated with the inner infrared nebula, but observations
of higher angular resolution are needed to spatially resolve this emission and establish its origin.

We have studied the star formation activity around G24 and discovered seven YSOs with spectral characteristics of 
protostars, projected on the molecular cloud that interestingly coincides with the IRDC 024.789+0.633.
Thus, this region harbors the typical components commonly found in the vicinity of star forming regions, 
namely an IRDC, several solar masses of molecular gas, and embedded YSO candidates. 
We propose that the birth of these young objects might have been triggered by the expanding stellar wind bubble. 

From the study of the distribution of the molecular gas in a large region toward G24, we suggest a link
between the origin of the LBV star and the progenitor star of the nearby SNR G24.7+0.6. This SNR is 
also interacting with the neighboring molecular gas and shows star-forming activity in its surroundings.
We propose a scenario where both massive stars belong to a first generation of stars that formed from the same 
natal MC, pointing to the possibility of finding other massive stars with a common origin.
The strong stellar winds of these massive stars might have triggered the formation of a second generation of stars, which
now appear as protostars deeply embedded in the molecular gas.

\section*{Acknowledgments}

We wish to thank the anonymous referee whose comments and suggestions have helped to improve the paper.
A.P. is a doctoral fellow of CONICET, Argentina. S.P. and E.G. are members of 
the {\sl Carrera del investigador cient\'\i fico} of CONICET, Argentina.
This research was partially supported by Argentina Grants awarded by CONICET, ANPCYT and 
University of Buenos Aires (UBACYT).

\bibliographystyle{aa}  
\bibliography{biblio}
\IfFileExists{\jobname.bbl}{}
{\typeout{}
\typeout{****************************************************}
\typeout{****************************************************}
\typeout{** Please run "bibtex \jobname" to obtain}
\typeout{** the bibliography and then re-run LaTeX}
\typeout{** twice to fix the references!}
\typeout{****************************************************}
\typeout{****************************************************}
\typeout{}
}

\label{lastpage}

\end{document}